\documentstyle[11pt]{article}
\begin{document}
\renewcommand
\baselinestretch{1.0}
\textwidth 15.0 true cm
\textheight 22.0 true cm
\headheight 0 cm
\headsep 0 cm
\topmargin 0.4 true in
\oddsidemargin 0.25 true in

\title{Direct Monte Carlo Measurement of the Surface \\
       Tension in Ising Models }
\author{Martin Hasenbusch \\
        Fachbereich Physik, Universit\"at  Kaiserslautern \\
        D-6750 Kaiserslautern, Germany}
\date{   }
   \date{\today\\
      Kaiserslautern Preprint KL--TH~16/92}
     \message{ surface tension, \today}
\maketitle
\begin{abstract}
I present a cluster Monte Carlo algorithm that gives direct access to
the interface free energy of Ising models. The basic idea is to simulate
an ensemble that consists of both configurations with periodic and with
antiperiodic boundary conditions. A  cluster algorithm is provided that
efficently updates this joint ensemble. The interface tension is
obtained from the ratio of configurations with periodic and antiperiodic
boundary conditions, respectively. The method is tested for the
3-dimensional Ising model.
\end{abstract}
\nopagebreak[4]
\newpage

\section{Introduction}

The interfaces of 2D and 3D Ising models at temperatures below the bulk
critical temperature $T_c$ have been studied as models of interfaces
separating coexisting phases of fluids. There are also
relations to lattice gauge theory: The surface tension of the 3D
Ising model is equal to the string  tension of the 3D $Z_2$ gauge model
which is dual to the 3D Ising  model.

While in the 2D case a number of exact results have been obtained, Monte
Carlo simulations play a major role in the study of 3D systems.
Recently a number of simulations employing various methods have
been performed to determine the surface tension of 3D and 4D Ising
models
 \cite{Meyer,MuensKles,BergHans,Sanie,HaPi,JanShen}, while in ref.
 \cite{Case} the string tension of the
 3D $Z_2$ gauge model is studied.

 As the temperature $T$ increases towards the critical temperature $T_c$,
  the
 reduced surface tension $\sigma = \tau / \beta $, where $\tau$ is the
 surface tension and $\beta$ the  inverse temperature,
  vanishes according to the scaling law
 \begin{equation}
 \sigma = \sigma_0 t^{\mu} ,
 \end{equation}
 where
 $t = (T_c - T)/T_c$,
  and $\sigma_0$ is the critical amplitude
  of the reduced interface tension.
  Widoms scaling law \cite{Widom2,Widom1}
 \begin{equation}
  \mu = (D-1) \nu
 \end{equation}
  relates the universal critical exponent $\mu$ to the critical exponent
 of the correlation length
 \begin{equation}
 \xi = \xi_0 t^{-\nu} .
 \end{equation}
In a recent Monte Carlo Renormalization Group study
of the 3D Ising model on a simple cubic lattice
\cite{Baillie} $\beta_c=0.221652(4)$ and $\nu=0.624(2)$ have been
obtained, while $\epsilon$-expansion predicts $\nu=0.630(2)$ \cite{eps}.
 The experimental \cite{Pegg,Gielen,Moldover,Chaar} value  for $\mu$ is
 $\mu = 1.26(1)$,
 consistent with Widoms scaling law.
  Ratios of critical amplitudes should also be universal due to the
  scaling hypothesis \cite{Fisk,Stauffer}.
  Experimental results for various binary systems are consistent with
  \begin{equation}
    R_+ = \sigma_0 (\xi_0^+)^2 = 0.386
  \end{equation}
  \cite{Chaar}, where $\xi_0^+$ is the critical correlation length
  amplitude in the high temperature phase.

  An  interesting question is the relation of the surface tension
  with the correlation length of a system with cylindrical geometry, i.e.\
   a system on a lattice with extension $L \times  L \times T $,
  where $T \gg L $.
  Recently, Borgs and Imbrie \cite{BoIm}
  gave an exact derivation of the finite size
  behaviour of the correlation length of discrete spin systems in a
  cylindrical geometry. They claim that for sufficiently large
  couplings the properties of the system are given by an effective
  1D model, where the diagonal parts of the transfer matrix are given
  by the free energies of the pure phases, while the off diagonal
  elements are determined by the surface tensions between the different
  phases.
  For the 3D Ising model this leads to the relation
\begin{equation}
  \xi_L = \exp(\sigma L^2) \, .
\end{equation}
 A semiclassical instanton calculation \cite{Mue3d} however
 predicts
\begin{equation}
  \xi_L = c \exp(\sigma L^2) \, ,
\end{equation}
 where $c$ depends on the temperature and is not equal to 1.

  In order to understand this discrepancy
  I compared the correlation length of an 1D Ising model with
\begin{equation}
    \label{beteff}
   2 \beta_{\rm eff} = F_s \,  ,
\end{equation}
   where $F_s$ is the reduced surface free energy,
   with the correlation length measured in ref.
  \cite{MuensKles} for 3D Ising cylinders.

  The correlation length $\xi$ of a 1D
  Ising model is given by
\begin{equation}
 \label{xieff}
 \xi = \frac{1}{\ln((1+v)/(1-v))} \, ,
\end{equation}
 where $v=\exp(-2\beta)$.
  For large $\beta$ one gets approximately
\begin{equation}
 \xi = \frac{1}{2v} \, .
\end{equation}

This paper is organized as follows. First I explain the model with
 periodic and antiperiodic boundary conditions. I discuss how one can
 get the surface
 tension from observables of a system which includes  the boundary
conditions as dynamical variables. Then I present a cluster algorithm
which is suitable for the simulation of such a system. Finally the
numerical results will be given and  compared with recent Monte Carlo
studies employing other methods.

\section{The Model}
 I consider a simple cubic lattice with extension $L$ in x- and
 y-direction and with extension T in z-direction.
 The uppermost layer of the lattice is regarded as the lower neighbor
 plane of the lowermost plane. An analog identification is done for
 the other two lattice directions.
The Ising model is defined by the Hamiltonian
\begin{equation}
  H(s,b.c.) = - \sum_{<ij>} J_{<ij>} s_i s_j .
\end{equation}
 When periodic (p.) boundary
 conditions (b.c.) are employed, then $J_{<ij>}=1$
for all nearest neighbor pairs. When      antiperiodic (a.p.) boundary
conditions are employed, then
           $J_{<ij>}=-1$ for bonds $<ij>$ connecting the lowermost and
 uppermost plane of the
 lattice, while all other nearest neighbor pairs keep $J_{<ij>} = 1$.

\section{The Surface Tension}

  I consider a system that allows both periodic and antiperiodic
boundary conditions. The partition function of this system is given by
\begin{equation}
 Z = \sum_{b.c.} \sum_{s_i=\pm 1} \exp(-\beta H(s , b.c.)) \, .
\end{equation}
 The fraction of configurations with antiperiodic boundary conditions
 is given by the ratio  $Z_{a.p.} / Z$ \, ,
\begin{eqnarray}
  \frac{Z_{a.p.}}{Z} &=& \frac{\sum_{s_i=\pm 1} \exp(-\beta H(s,a.p.))} {Z}
 \nonumber \\
 &=& \frac{\sum_{b.c.} \sum_{s_i=\pm 1} \exp(-\beta H(s , b.c.))
  \delta_{b.c.,a.p.}} {Z}  \nonumber\\ &=& <\delta_{b.c.,a.p.}> \, .
\end{eqnarray}
  An analogous result can be found for periodic boundary conditions.
 Now we can express the ratio $ Z_{a.p.} / Z_{p.} $ as a ratio of
 observables in this system.
\begin{equation}
 \frac{Z_{a.p.}}{Z_{p.}} = \frac{ \frac{Z_{a.p.}}{Z}}
                               { \frac{Z_{p.}}{Z}}
                         =\frac{<\delta_{b.c.,a.p.}>}
                              {<\delta_{b.c.,p.}>} \, .
\end{equation}
 In the case of a surface with fixed position,
 the surface free energy is given by
\begin{equation}
 F_s = F_{a.p.} - F_{p.} = \ln Z_{p.} - \ln Z_{a.p.} = -\ln \frac{Z_{a.p.}}
{Z_{p.}} \, ,
\end{equation}
 where $F_{p.}$ and $F_{a.p.}$ are the reduced
   free energies of the systems with
 periodic and antiperiodic boundary conditions, respectively.
 If we assume that there is no interface in the
 system with perodic boundary conditions and exactly one in the case of
 antiperiodic boundary conditions, we can take into account
    the entropy due to the
 free position of the interface in $T$ direction by
 adding $\ln T$,
\begin{equation}
 F_s = F_{a.p.} - F_{p.} + \ln T \, .
\end{equation}
We get a more appropriate description for finite systems
if we take into account the
occurance of several interfaces, an even number for periodic and an
odd number for antiperiodic boundary conditions. If we furthermore
assume that these interfaces do not interact we get an improved
  expression
\begin{equation}
 \tanh(\exp(-F_{s,i} + \ln T)) = \frac{Z_{a.p.}}{Z_{p.}}
\end{equation}
  for the surface free energy.
 If we resolve this equation with respect to $F_{s,i}$ we get
\begin{equation}
 \label{several}
  F_{s,i} =  \ln T - \ln ( \frac{1}{2} \ln (\frac {1 + Z_{a.p.}/Z_{p.}}
                                            {1 - Z_{a.p.}/Z_{p.}})) \, .
\end{equation}
\section{The Algorithm}
I shall now describe an efficent algorithm to update the above explained
 system, where the
type of boundary condition is a random variable.
 The simplest way to alter the boundary conditions is to propose a
 change of the coupling $J_{<ij>}$ of sites in the uppermost plane  with
 sites in the lowermost plane from $1$ to $-1$ or vice versa in a
 single  Metropolis step.
 With high probability most of the spins $s_i$ and $s_j$
 have the same
 sign in the case of
  periodic boundary conditions and different sign in the case of
  antiperiodic
 boundary conditions.
 Hence the  acceptance rate of such a Metropolis
step will be extremely small.
 This simple algorithm does not take into account that the physical
 interface can be built anywhere in the system and, what is even more
 important, that the interface wildly fluctuates close to the critical
 point.

The cluster algorithm is the natural candidate to find the physical
 surface structure.
  First one  goes through the lattice and deletes the bonds with the
                         standard  probability \cite{Wang,Wolff}
 \begin{equation}
  p_d = \exp(-\beta (1+J_{<ij>} s_i s_j)) .
 \end{equation}
  Then one searches for a sheet of deleted bonds that completely cuts
  the lattice in z-direction. If there is such a sheet the spins
  from the lowermost plane up to this sheet are flipped.
  This is a valid update, since the bonds in the sheet are deleted and
  the value of $J_{<ij>} s_i s_j$ for $i$ in the lowermost and $j$ in
  the uppermost plane is not changed when we alter the sign of
  $J_{<i,j>}$ and $s_i$.

  In my simulations I alternate this boundary flip update with a
  standard single cluster update \cite{Wolff}.
\section{Numerical Results}
I simulated the 3D Ising model on a simple cubic lattice
          with boundary conditions as dynamical
variables at $\beta=0.223, 0.224, 0.2255, 0.2275, 0.2327$ and $0.2391$.
 For most of the simulations lattices of size $ L \times L \times T $
 with $T = 3L$
 were used. In order to check for the $T$-dependence of the results
 at $\beta=0.2275$ also simulations with $T=L/2 , L , 2L$ were performed.
The statistics
of the simulations was 100000 times one single cluster update
   \cite{Wolff}       plus one
boundary flip update throughout.
I measured the  energy
 \begin{equation}
  E = \sum_{<ij>} J_{<ij>} s_i s_j ,
 \end{equation}
 the magnetization
 \begin{equation}
  m = \frac{1}{L^2 \times T} \sum_i s_i
 \end{equation}
   and the type of boundary condition (b.c.) after each pair of
single cluster plus boundary flip update.  These data are
  used to calculate
the energy density of the system with periodic boundary conditions
 \begin{equation}
 E_p = \frac {1}{L^2 \times T}
       \frac{\sum_{n=n_0}^{N} E \delta_{b.c.,p.} }
           {\sum_{n=n_0}^{N}  \delta_{b.c.,p.} } ,
 \end{equation}
  where $n$ labels the measurements, and $N$ is the number of measurements.
     The mean square magnetization of the system with
periodic boundary conditions is
 \begin{equation}
 <m^2> = \frac{\sum_{n=n_0}^{N} m^2 \delta_{b.c.,p.} }
           {\sum_{n=n_0}^{N}  \delta_{b.c.,p.} } ,
 \end{equation}
   and   the surface energy density
   \begin{equation}
 SE = \frac {1}{L^2} \Bigl(
       \frac{\sum_{n=n_0}^{N} E \delta_{b.c.,p.} }
           {\sum_{n=n_0}^{N}  \delta_{b.c.,p.} } -
       \frac{\sum_{n=n_0}^{N} E \delta_{b.c.,a.p.} }
           {\sum_{n=n_0}^{N}  \delta_{b.c.,a.p.} }
           \Bigr) .
 \end{equation}
    The results for these quantities are summarized in table 1.
   For parameters where the fraction of configurations with
antiperiodic  boundary conditions is large the value for the surface
energy is not reliable, since many of the configurations contain more
 than the minimal number of  interfaces. A strong
 dependence of $SE$ on $L$ is visible.

  Starting from the fraction of  configurations with
 antiperiodic boundary conditions $<\delta_{b.c.,a.p.}>$
 the reduced surface free energies $F_s$ and $F_{s,i}$
  are determined following eqn.\ (15) and (17), respectively.
  The  results are summarized in table 2.
   For $F_s \geq 6  $ the difference between the two definitions
   $F_s$ and $F_{s,i}$
  of the  surface energy  is smaller than the statistical errors.
  At $\beta = 0.2275$ I investigated the dependence of the
  surface free energy  on $T$. One can observe that
  $F_{s,i} $ remains constant within errorbars for $L=10, 12$ and $14$
  starting from $ T = L$. $ T = 3L$ seems to be safe not to spoil the
      results.

  Using  $F_{s,i}$ I calculated the inverse correlation length
   of an 1D Ising model
  with $2 \beta_{\rm eff} =  F_{s,i}$ following eq.\ (\ref{xieff}).
  The results
  which are given in table 2  can be compared with the direct
  measurement of the mass of a 3D Ising model on a cylindrical lattice
  at $\beta = 0.2275,0.2327$ and
  $0.2391$ of ref.\ \cite{MuensKles}.
  The numbers they give for $E_{0a}$ in their table 1 are consistent
  with my results for the masses of the effective 1D Ising model within
  errorbars.

  Similar to the surface energy the values of
   $ F_s / L^2 $ and $F_{s,i}/L^2$ which I give in table 2
    displays a strong dependence on the lattice
 size. It seems difficult to extract the infinite $L$
  limit of the surface
  tension from these numbers.
        Motivated by free field theory (in ref. \cite{HaPi}
    we demonstrate that the long range properties of an interface
    in the rough phase of a 3D Ising model is well described by a
    massless free field theory),
  I tried to fit the
  surface free energy according to the Ansatz
   \begin{equation}
  F_{s,i} = C + \sigma L^2 \, .
   \end{equation}
   It turned out that  the data fit very well to   this
   Ansatz.
  The results of the  fits are given in table 3.

  Starting from the $\sigma$'s given in table 3 I did several fits
  to test the scaling  law $\sigma = \sigma_0 t^{\mu} $.
   I used two different
  definitions for  the reduced temperature,
    $t_1 =   (\beta-\beta_c)/\beta_c $ and
    $t_2 = (T_c-T)/T_c$. In both cases I used $\beta_c = 0.221652$
    given in ref. \cite{Baillie}.
    Remember that $t_1$ and $t_2$ are equivalent in the first
    order of a Taylor series around $T_c$.
   The results are given in table 4 and table 5.
   One can observe that is neccesary to go even closer to the critical
   temperature to overcome the ambiguity  in the definition
   of the reduced temperature $t$. Taking into account this systematic
   errors      I get as an estimate for the critical exponent
   $\mu=1.24(3)$.

    In order to get a better estimate for the critical amplitude of
    the surface tension $\sigma_0$
    I  used the results of ref.
    \cite{Baillie,eps} for $\nu$ combined with
    the scaling relation $\mu = 2\nu$ and determined
   \begin{equation}
      \sigma_0 = \sigma t^{-\mu}
    \end{equation}
    from single measurements of $\sigma$.
    The results are given in table 6.
   Taking into account the uncertainty in the value of $\nu$
   a final estimate  $\sigma_0 = 1.5 \pm 0.1$ seems reasonable.
   Using the estimate $\xi_0^+ = 0.4783 \pm 0.0004$ of ref.
   \cite{TaFi} I get $R_+ = 0.34(2)$.   Taking into account the
    deviation from the mean value
   of the results for the various binary alloys quoted in ref.\
    \cite{Chaar}
    my
   result is well consistent with experiment and most of the recent
   Monte Carlo simulations \cite{Mon,MuensKles,BergHans,Case}.
    Since I have surface tensions for more $\beta$ values and $\beta$'s
   closer to the phase transition as the references quoted above
   I improved the control on finite
    $t$ effects.
   One should mention  that ealier results of Monte Carlo simulation
   \cite{Binder}   and  analytic calculations \cite{Brezin}  were
    about 30\% below the experimental value.

   Let me finally comment on the performance of the algorithm.
   The autocorrelation times were of order 1 in units of the
  combined single cluster plus boundary flip update for all simulations
  quoted above. The simulation of the largest system
    ($36 \times 36 \times 108$) took 84h on an IBM risc  station 6000.
   The drawback of the method is its limitation to small surface
   free energies. For $F_s > 9$ the fraction of configurations with
   antiperiodic boundary conditions becomes smaller than 1\% and hence
   it is hard to get a sufficient statistic of configurations with
   antiperiodic boundary conditions.
   A solution of this problem might be found in a combination with
   multicanonical methods.  But the most naive proposal of this kind,
    just to introduce a chemical potential that makes the antiperiodic
    boundary conditions more probable, fails. The flip from periodic
    boundary conditions to antiperiodic boundary conditions is allowed
    only if there is a sheet of deleted bonds in the system that
    cuts the
    lattice. The chemical potential just forces the system to stay
    longer with   antiperiodic boundary conditions after such a flip.
     Hence the statistics of boundary flips  is even reduced.
%
%

 \section{Conclusion}
  I  presented an effective  method to determine the surface tension of
  Ising systems. It should also be applicable to other
  discrete spin models.
   The method allowed to obtain the surface tension very close
  ($T=0.994 \, T_c$) to the critical temperature with a high accuracy.
  The mass of the cylindrical 3D Ising system due to tunneling turned
  out to be given to a very good accuracy by the mass of an 1D Ising
  model with $2 \beta_{\rm eff} =$ surface free energy, which is
 consistent with the prediction of ref. \cite{BoIm}.
  But the finite size behavior of the surface free energy of the rough
  interface  is well described by  $F_s = C + \sigma L^2$ leading to
  the prefactor predicted in ref.\ \cite{Mue3d}.

 \section{Acknowledgement}
   I thank S. Meyer,
   G. M\"unster and K. Pinn for useful discussions. This
   work was supported by the Deutsche Forschungs Gemeinschaft through
   Grant No. Me 567/5-3 and the German-Israeli Foundation of
   Scientific Research.
   The
   simulations were performed on the IBM RISC system/6000  of the
   Regionales Hochschulrechenzentrum Kaiserslautern (RHRK).

 \begin{table}
 \caption{Data for 3-D Ising cylinders of size $ L^2 \times T $
 at $\beta = 0.223$, $0.224$, $0.2255$, $0.2275$, $0.2327$ and
 $0.2391$. $E_p$ denotes the energy density of the system with
 periodic boundary conditions, $<m^2>$ is the expectation
 value of the square magnetization of the system with
  periodic boundary conditions.
 $ES$ is the difference of the energy with periodic and the energy with
 antiperiodic boundary conditions divided by the area $L^2$.
 $<\delta_{b.c.,a.p.}> $
            gives the fraction of configurations
    with antiperiodic boundary conditions.}
 \label{gaee}
  \begin{center}
   \begin{tabular}{|r|r|l|l|l|l|}
   \hline
   L & $ T $& $E_p$ & $<m^2>$ & $ES$ &
   $<\delta_{b.c.,a.p.}>$ \\
   \hline
   $\beta$=0.2391 \\
   \hline
  4 & 12 & 0.5219(9) & 0.3586(14) &  1.364(41) &
           0.4388(8) \\
   \hline
  6 & 18 & 0.5435(5) &  0.4187(9) & 2.607(40) &
           0.2948(12) \\
   \hline
  8 & 24 & 0.5524(3) & 0.4432(5) &  3.247(41) &
           0.1075(10) \\
   \hline
  10 & 10 & 0.5541(4) &   0.4497(6) & 3.378(57)   &
            0.0066(3) \\
  10 & 30 & 0.5536(2) & 0.4460(3) & 3.433(66) &
            0.0192(5) \\
   \hline
   $\beta$=0.2327 \\
   \hline
  8 & 24 & 0.4861(4) & 0.3310(8) &  2.498(42) &
           0.2833(13) \\
   \hline
  10 &  30 & 0.4905(3) & 0.3459(5) &  2.986(44) &
           0.1359(13) \\
   \hline
  12 & 36 & 0.4917(2) & 0.3487(4) &  3.166(55) &
            0.0445(7) \\
   \hline
  14 & 42 & 0.4921(2) & 0.3493(3) & 3.223(81)  &
            0.0100(3) \\
   \hline
   $\beta$=0.2275 \\
   \hline
 10 & 5 &  0.4219(8) & 0.2381(13) &  1.742(15) &
              0.0782(11) \\
 10 & 10 & 0.4284(6) & 0.2518(10) & 2.167(24) &
              0.1775(16) \\
 10 & 20 & 0.4250(5) & 0.2334(9) & 2.021(39) &
              0.2908(15) \\
 10 & 30 & 0.4231(4) &  0.2176(9) &  1.810(50) &
             0.3633(13)  \\
   \hline
 12 & 6 & 0.4227(7) &  0.2325(12) & 1.877(17) &
             0.0551(10) \\
 12 & 12 & 0.4288(5) &  0.2472(9) &  2.383(29) &
             0.1179(16) \\
  12 & 24 & 0.4272(4)  & 0.2377(7) &  2.360(39) &
             0.2061(17) \\
  12 & 36 & 0.4268(3) &  0.2301(7) &  2.269(49) &
             0.2741(16) \\
   \hline
  14 & 14 & 0.4296(4) &  0.2461(8) &  2.604(30) &
             0.0683(13) \\
  14 & 28 & 0.4294(3) &  0.2415(6) &  2.590(40) &
             0.1232(15) \\
  14 & 42 & 0.4293(3) &  0.2388(5) &  2.605(47) &
             0.1698(16) \\
   \hline
  16 & 48 & 0.4301(2) &  0.2418(4) &  2.706(48) &
             0.0860(12) \\
   \hline
  18 & 54 & 0.4302(2) & 0.2422(3) &  2.836(62) &
             0.0378(8) \\
   \hline
   \end{tabular}
  \end{center}
 \end{table}
 \begin{table}
  \begin{center}
   \begin{tabular}{|r|r|l|l|l|l|}
   \hline
   $L$ & $ T $& $E_p$ & $<m^2>$ & $ES$ &
   $<\delta_{b.c.,a.p.}>$ \\
   \hline
   $\beta$=0.2255 \\
   \hline
  14 & 42 & 0.3982(3) & 0.1734(7) & 1.805(53) &
            .3281(16) \\
   \hline
  16 & 48 & 0.4001(3) & 0.1808(6) & 2.121(51) &
            0.2540(18) \\
   \hline
  18 & 54 & 0.4012(2) &  0.1852(5) &  2.310(46) &
            0.1752(16) \\
   \hline
  20 & 60 & 0.4018(2) &  0.1872(4) & 2.426(47) &
            0.1076(14) \\
   \hline
  24 & 72 & 0.4022(1) & 0.1885(3) &  2.490(68)  &
            0.0284(7) \\
   \hline
   $\beta$=0.224 \\
   \hline
  14 & 42 &  0.3731(3) & 0.1167(8) & 1.014(56)&
            0.4252(12) \\
   \hline
   18 & 54 & 0.3750(3) & 0.1247(7) & 1.467(56) &
           0.3553(17) \\
   \hline
   24 & 72 & 0.3778(2) & 0.1362(4) & 1.982(50) &
           0.1884(20) \\
   \hline
   30 & 90 & 0.3788(1) & 0.1395(3) & 2.312(65) &
           0.0575(12) \\
   \hline
   $\beta$=0.223 \\
   \hline
   8 & 24 & 0.3648(5) & 0.1077(9) &   0.380(52) &
            0.4821(5) \\
   \hline
  12 & 36 & 0.3588(4) &  0.0879(8) & 0.469(58)  &
            0.4720(6) \\
   \hline
  18 & 54 & 0.3572(3) &  0.0812(8) &  0.744(72) &
            0.4423(13) \\
   \hline
  24 & 72 & 0.3586(2) &   0.0867(6) & 1.186(59) &
            0.3754(19) \\
   \hline
  30 & 90 & 0.3600(2) &  0.0938(4) & 1.605(58) &
            0.2677(23) \\
   \hline
  36 & 108 & 0.3608(1) &  0.0975(3) & 1.845(56) &
            0.1437(21) \\
   \hline
   \end{tabular}
  \end{center}
 \end{table}
 \begin{table}
 \caption{Results for the surface tension
  and the mass $(mass_{1d})$ of an effective Ising model with
  $2 \beta_{\rm eff}$ = $F_{s,i}$ are given.
 $ F_s$ and $F_{s,i}$ are explained in the text.
     }
 \label{gaeq}
  \begin{center}
   \begin{tabular}{|r|r|l|l|l|l|l|}
   \hline
  $ L$ & $T$ & $F_s$ & $F_s/L^2$ & $F_{s,i}$
    & $F_{s,i}/L^2$ &
  $ mass_{1d}$ \\
   \hline
   $\beta$=0.2391 \\
   \hline
  4 & 12 &
             2.731(3) &  0.1707(2) &
             2.436(6) &  0.1523(4) & 0.1755(10)   \\
   \hline
  6 & 18 &
            3.763(6) &   0.1045(2) &
            3.699(7) &   0.1028(2) & 0.0495(3)    \\
   \hline
  8 & 24 &
             5.294(11) & 0.0827(2) &
             5.289(11) &  0.0827(2) & 0.01010(11)   \\
   \hline
  10 & 10 &
            7.311(40) &   0.0731(4) &
            7.311(40) &   0.0731(4) & 0.00134(5) \\
  10 & 30 &
            7.335(24) &   0.0734(3) &
            7.335(24) &   0.0734(3) & 0.00130(3) \\
   \hline
   $\beta$=0.2327 \\
   \hline
  8 & 24 &
             4.106(7) &  0.06416(10)         &
             4.050(7) &  0.06328(12)      & 0.0348(2) \\
   \hline
  10 & 30 &
             5.251(11)  &  0.05251(11)   &
             5.243(11)  &  0.05243(11)  & 0.01057(12)   \\
   \hline
  12 & 36 &
            6.649(17) &  0.04618(12) &
            6.649(17) &  0.04617(12) &  0.00259(4) \\
   \hline
 14 & 42 &
             8.330(33) &  0.04250(17)  &
             8.330(33) &  0.04250(17)  & 0.00048(2)  \\
   \hline
   $\beta$=0.2275 \\
   \hline
 10 & 5 &
            4.076(15) & 0.04076(15) &
              4.074(15) & 0.04074(15) & 0.0340(5) \\
 10 & 10  &
          3.836(11) & 0.03836(11) &
              3.820(11) & 0.03820(11) & 0.0439(5) \\
 10 & 20
         & 3.887(7) &   0.03887(7) &
              3.827(8) &  0.03827(8) & 0.0436(4) \\
 10 & 30 &
           3.962(6)  &  0.03962(6) &
             3.834(8)  &  0.03834(7) & 0.0433(4) \\
   \hline
 12 & 6
        & 4.634(19) &  0.03218(13) &
             4.633(19) &  0.03217(13) & 0.0195(4) \\
 12 & 12
         & 4.497(15) &  0.03123(10) &
             4.491(15) &  0.03119(10) &  0.0224(3) \\
  12 & 24
         &  4.527(10) &  0.03144(7) &
             4.504(11) &  0.03128(7) & 0.0221(2) \\
  12 & 36
          & 4.557(8)  &  0.03165(6) &
             4.507(9) &   0.03130(6) &  0.0221(2) \\
   \hline
  14 & 14 &
              5.252(20) &  0.02679(10) &

             5.250(20) &  0.02679(10) & 0.0105(2) \\
  14 & 28 &
               5.294(13) &  0.02701(7) &
             5.288(14) &  0.02698(7) & 0.01010(14) \\
  14 & 42 &
               5.325(11) &  0.02717(6) &
             5.311(11) &  0.02710(6) & 0.00987(11) \\
   \hline
  16 & 48 &
               6.234(15) &  0.02435(6) &
             6.231(15) &  0.02434(6) & 0.00393(6) \\
   \hline
  18 & 54 &
              7.225(21) &  0.02230(6) &
             7.225(21) &  0.02230(6) & 0.00146(3) \\
   \hline
   \end{tabular}
  \end{center}
 \end{table}
 \begin{table}
  \begin{center}
   \begin{tabular}{|r|r|l|l|l|l|l|}
   \hline
   L & $T$ & $F_s$ & $F_s/L^2$ & $F_{s,i}$
    & $F_{s,i}/L^2$ &
  $ mass_{1d}$ \\
   \hline
   $\beta$=0.2255 \\
   \hline
 14 & 42 &
             4.455(7) &  0.02273(4) &
             4.365(9) &  0.02227(5) & 0.0254(2)   \\
   \hline
 16 & 48 &
             4.949(9) & 0.01933(4) &
             4.908(10) & 0.01917(4) & 0.0148(1)    \\
   \hline
 18 & 54 &
             5.537(11) & 0.01709(4) &
             5.522(11) & 0.01704(4) & 0.00800(9)      \\
   \hline
 20 & 60 &
             6.210(15) & 0.01552(4) &
             6.205(15) & 0.01551(4) & 0.00404(6)   \\
   \hline
 24 & 72 &
             7.808(25) & 0.01356(4) &
             7.807(25) & 0.01355(4) & 0.00081(2)    \\
   \hline
   $\beta$=0.224 \\
   \hline
  14 & 42 &
             4.039(5) & 0.02061(2) &
             3.789(8) & 0.01933(4) & 0.0452(3)   \\
   \hline
   18 & 54 &
               4.585(7) & 0.01415(2) &
               4.467(9) & 0.01379(3) & 0.0230(2)   \\
   \hline
   24 & 72 &
               5.737(13) & 0.00996(2) &
               5.719(13) & 0.00993(2) & 0.00657(8)  \\
   \hline
   30 & 90 &
              7.298(22) & 0.00811(2) &
              7.296(22) & 0.00811(2) &  0.00136(3)  \\
   \hline
   $\beta$=0.223 \\
   \hline
        8 & 24 &
             3.250(2) & 0.05078(3) &
             2.669(9) & 0.04170(13) & 0.139(1)  \\
   \hline
       12 & 36 &
             3.696(3) & 0.02566(2) &
             3.218(8) & 0.02235(6) & 0.0801(6)   \\
   \hline
       18 & 54 &
             4.221(6) & .01303(2) &
             3.912(11) & 0.01207(3) &  0.0400(4)  \\
   \hline
       24 & 72 &
             4.786(8) & 0.00831(1) &
             4.641(11) & 0.00806(2) & 0.0193(2) \\
   \hline
       30 & 90 &
             5.506(12) & 0.00612(1)  &
             5.459(13) & 0.00607(1) &  0.00852(10) \\
   \hline
       36 & 108 &
            6.467(17) & 0.00500(1) &
            6.457(17) & 0.00498(1) &  0.00314(5)  \\
   \hline
   \end{tabular}
  \end{center}
 \end{table}
 \begin{table}
 \caption{Results of fits of the form $F_{s,i} = C + \sigma L^2$
   are given. Only values from the largest $T$ are included in the fits.
 $\chi^2$/d.o.f. denotes the square deviation per degrees of freedom.}
 \label{gweq}
  \begin{center}
   \begin{tabular}{|c|c|c|c|c|}
  \hline
    $\beta$ &$L$'s used& $C$ & $\sigma$ & $\chi^2$/d.o.f. \\
   \hline
      0.2391 &6,8,10 & 1.65(2) & 0.0568(3)& 0.002 \\
   \hline
      0.2327 &8,10,12,14  &1.97(2) &0.0325(2)&1.74 \\
   \hline
      0.2275 &12,14,16,18 &2.32(2) & 0.01521(11) &1.69 \\
   \hline
      0.2255  &14,16,18,20,24 &2.59(2) &0.00904(6)& 0.08    \\
   \hline
      0.224  & 18,24,30 & 2.87(2) & 0.00492(4) & 0.63 \\
   \hline
      0.223 &24,30,36 & 3.19(2) & 0.00252(3) & 0.002 \\
   \hline
   \end{tabular}
  \end{center}
 \end{table}
 \newpage

 \begin{table}
 \caption{Fits of the form $\sigma = \sigma_0 t_1^{\mu}$, where
 $t_1 = (\beta - \beta_c)/\beta_c$. The labels $1, 2, 3, 4, 5$ and
  $6$ correspond to $\beta = 0.2391, 0.2327, 0.2275, 0.2255, 0.224$
   and $0.223$, respectively.}
 \label{geww}
  \begin{center}
   \begin{tabular}{|c|c|c|c|}
  \hline
  input & $\mu$ & $\sigma_0$ & $\chi^2$/d.o.f. \\
\hline
   1,2,3,4,5,6 & 1.217(4) & 1.25(2) & 0.89 \\
     2,3,4,5,6 & 1.218(5) & 1.26(3) & 1.15 \\
       3,4,5,6 & 1.228(8) & 1.32(4) & 0.58 \\
         4,5,6 & 1.220(12)& 1.27(7) & 0.42 \\
     1,2,3,4,5 & 1.217(4) & 1.25(2) & 0.89 \\
       2,3,4,5 & 1.218(6) & 1.26(3) & 1.15 \\
         3,4,5 & 1.237(12) & 1.36(7) & 0.09 \\
   \hline
   \end{tabular}
  \end{center}
 \end{table}
 \begin{table}
 \caption{Fits of the form $\sigma = \sigma_0 t_2^{\mu}$, where
 $t_2 = (T_c - T)/T_c$. The labels $1, 2, 3, 4, 5$ and
  $6$ correspond to $\beta = 0.2391, 0.2327, 0.2275, 0.2255, 0.224$
   and $0.223$, respectively.}
 \label{gewe}
  \begin{center}
   \begin{tabular}{|c|c|c|c|}
  \hline
  input & $\mu$ & $\sigma_0$ &$\chi^2$/d.o.f. \\
\hline
   1,2,3,4,5,6 & 1.256(4)  & 1.51(2) & 2.5 \\
     2,3,4,5,6 & 1.246(5)  & 1.45(3) & 0.8 \\
       3,4,5,6 & 1.246(9)  & 1.45(5) & 1.2 \\
         4,5,6 & 1.234(13) & 1.37(8) & 0.7 \\
     1,2,3,4,5 & 1.260(4)  & 1.53(2) & 1.6 \\
       2,3,4,5 & 1.250(6)  & 1.47(3) & 0.4 \\
         3,4,5 & 1.258(12) & 1.52(8) & 0.3 \\
   \hline
   \end{tabular}
  \end{center}
 \end{table}
 \begin{table}
 \caption{Results for $\sigma_0 = \sigma t_1^{-\mu}$
           and   $\sigma_0 = \sigma t_1^{-\mu}$ using the value of
           single measurements for $\sigma$ and given $\mu = 1.248$
  and $1.26$.
}
 \label{gwww}
  \begin{center}
   \begin{tabular}{|c|c|c|}
  \hline
 $\beta$ & $\sigma_0$, $t_1$ & $\sigma_0$, $t_2$ \\
\hline
$\mu$=1.248  \\
\hline
0.2391 & 1.355(7) &  1.490(8) \\
0.2327 & 1.372(8) &  1.458(9) \\
0.2275 & 1.420(10) &  1.467(11) \\
0.2255 & 1.423(9) &  1.454(10) \\
0.2240 & 1.435(12) &  1.454(12) \\
0.2230 & 1.469(18) &  1.480(18)  \\
\hline
 $\mu$=1.26  \\
\hline
0.2391 & 1.397(7) &  1.537(8)  \\
0.2327 & 1.422(9) &  1.512(9)  \\
0.2275 & 1.483(11) &  1.533(11) \\
0.2255 & 1.494(10) &  1.527(10)  \\
0.2240 & 1.515(12) &  1.535(13) \\
0.2230 & 1.562(19) &  1.574(19) \\
\hline
   \end{tabular}
  \end{center}
 \end{table}

\newpage

\end{document}